\documentclass[aps,prl,twocolumn,superscriptaddress]{revtex4}
\usepackage{amsmath,amssymb,braket,graphicx,bm}

\begin{document}
\title{Topological Corner States in Graphene by Bulk and Edge Engineering}
\author{Junjie Zeng}
\thanks{These two authors contributed equally.}
\affiliation{CAS Key Laboratory of Strongly-Coupled Quantum Matter Physics, and Department of Physics, University of Science and Technology of China, Hefei, Anhui 230026, China}	
\author{Chen Chen}
\thanks{These two authors contributed equally.}
\affiliation{International Center for Quantum and Molecular Structures, Materials Genome Institute, Physics Department, and Shanghai Key Laboratory of High Temperature Superconductors, Shanghai University, Shanghai 200444, China}
\author{Yafei Ren}
\affiliation{Department of Materials Science and Engineering, University of Washington, Seattle, Washington 98195, USA}
\author{Zheng Liu}
\affiliation{CAS Key Laboratory of Strongly-Coupled Quantum Matter Physics, and Department of Physics, University of Science and Technology of China, Hefei, Anhui 230026, China}	
\author{Wei Ren}
\email[Corresponding author:~]{renwei@shu.edu.cn}
\affiliation{International Center for Quantum and Molecular Structures, Materials Genome Institute, Physics Department, and Shanghai Key Laboratory of High Temperature Superconductors, Shanghai University, Shanghai 200444, China}
\author{Zhenhua Qiao}
\email[Corresponding author:~]{qiao@ustc.edu.cn}
\affiliation{CAS Key Laboratory of Strongly-Coupled Quantum Matter Physics, and Department of Physics, University of Science and Technology of China, Hefei, Anhui 230026, China}	
\affiliation{ICQD, Hefei National Laboratory for Physical Sciences at Microscale, University of Science and Technology of China, Hefei, Anhui 230026, China}

\begin{abstract}
  Two-dimensional higher-order topology is usually studied in (nearly) particle-hole symmetric models, so that an edge gap can be opened within the bulk one. But more often deviates the edge anticrossing even into the bulk, where corner states are difficult to pinpoint. We address this problem in a graphene-based $\mathbb{Z}_2$ topological insulator with spin-orbit coupling and in-plane magnetization both originating from substrates through a Slater-Koster multi-orbital model. The gapless helical edge modes cross inside the bulk, where is also located the magnetization-induced edge gap. After demonstrating its second-order nontriviality in bulk topology by a series of evidence, we show that a difference in bulk-edge onsite energy can adiabatically tune the position of the crossing/anticrossing of the edge modes to be inside the bulk gap. This can help unambiguously identify two pairs of topological corner states with nonvanishing energy degeneracy for a rhombic flake. We further find that the obtuse-angle pair is more stable than the acute-angle one. These results not only suggest an accessible way to ``find" topological corner states, but also provide a higher-order topological version of ``bulk-boundary correspondence".
\end{abstract}
\date{\today}
\maketitle

\textit{Introduction---.}
It has been well established so far that topologically nontrivial phases in crystalline solid materials can be understood in a unified and consolidated mathematical picture in terms of fiber bundle, constructed from Bloch wavefunctions over the first Brillouin zone as its base space, bearing a geometrical structure that is not globally direct-product decomposable (aka, with a global twisting of some kind)~\cite{Nakahara2003,Fruchart2013CRP,Zhao2013a,Kaufmann2016RMP,Cayssol2021JPM}. Typical examples are the quantum anomalous Hall effect~\cite{Liu2016ARCMP,Chang2013S,Qiao2010PRB} and quantum spin Hall effect~\cite{Kane2005PRL,Bernevig2006PRL,Koenig2007S}, whose corresponding Bloch bundles are respectively categorized by Chern class~\cite{Thouless1982PRL,Simons1983PRL} and $ \mathbb{Z}_2 $ class~\cite{Kane2005PRLa,Fu2007PRB,Moore2007PRB}, contributing therefore to their appellations as Chern insulator and $\mathbb{Z}_2$ topological insulator. On the other hand, another relevant and intensively reiterated conception is the ``bulk-boundary correspondence", because of not only its concreteness to comprehend, but also its prediction of perfect conducting channels with potential for exploitation. Though efforts have been devoted in its formulation~\cite{Qi2006PRB,Teo2010PRB,Fukui2012JPSJ,Graf2013CMP,Rhim2018PRB,Max2019PhD,Alase2019PhD,Silveirinha2019PRX,Yao2018PRL}, in contrast, it still can hardly be considered beyond a conjecture with respect to the whole research field of topological physics. One of its statements may read: a system accommodating a nontrivial bulk phase has its gapless representatives on its one-less-dimension boundaries if it changes to a finite geometry. It works, for example, for the aforementioned Chern insulator and $\mathbb{Z}_2$ topological insulator with manifestation as gapless chiral and helical edge states, respectively. However, this relationship turns out to be faced with the demanding for modification at least by two cases: non-Hermitian~\cite{Xiong2018JPC,Kunst2018PRL,Imura2019PRB,Gao2020PRL,Xiao2020NP} and higher-order topological systems~\cite{Kim2020LSA,Lee2020nQM,Park2019PRL,Otaki2019PRB,Schindler2018SA,Schindler2020JAP}. In the former case, one-by-one situations need corresponding generalizations, and in the latter case there are simply no gapless boundary states at all on boundaries with unit-lower dimensions. Among various attempts to rebuild it in a broader sense, as a consequence, e.g., one has to find smoking guns as one-dimensional hinge states~\cite{Tanaka2020PRB,Plekhanov2020PRR,Yue2019NP} and zero-dimensional corner states~\cite{Pelegri2019PRB,Luo2019PRL,Sheng2019PRL,Chen2020PRL,Ren2020PRL}, respectively, in three- and two-dimensional topological systems. But these processes are not always easy to accomplish and one of the difficulties is represented by that in two-dimensional systems the first-order gapless dispersion crosses outside of the bulk gap, making the ``corner states within gap in gap" strategy no longer applicable.

In this work, we resolve this problem by using monolayer graphene as a prototypical system, which is widely adopted as an ideal arena for various topologically nontrivial phases \cite{Haldane1988PRL,Kane2005PRL,Qiao2010PRB,Qiao2011PRL,Tse2011PRB,Qiao2012PRB,Qiao2013PRB,Deng2017PRB,Zeng2017PRB,Ren2020PRL,Liu2021PRR,Pan2022nCM}. Inspired by the ab initio research on a monolayer graphene system with bismuth ferrite as substrate \cite{Qiao2014PRL} (also see Sec. SI in Supplementary Materials~\cite{Zeng2021SM}) and the higher-order topology generating routine by opening gaps for gapless boundary modes \cite{Ren2020PRL,Xie2020NC}, we first confirm its bulk nontrivial topology despite a vanishing Chern number by both the gapless evolution of bulk Wannier charge center and the nonvanishing mirror-graded Zak phases ($ \gamma_\pm = \pi $), and then we find (i) the tunability of the edge states by an adiabatic onsite energy contrast between the bulk and the edge atoms, and (ii) an extra acute-angle pair of corner states degenerated on a different nonvanishing energy, which has a different origin from the obtuse-angle type. Those observations indicate that topological corner modes, as important evidence of higher-order nontrivial topology, may not be identifiable so directly as topological gapless edge ones, and therefore might facilitate a step forward in deepening our understanding in the guiding rule of ``bulk-boundary correspondence". Additionally, because of the similarity in mathematical descriptive formalism \cite{Tao1985JASA,Kushwaha1993PRL,Born1962,Li2009,Deymier2013,Khelif2016,Joannopoulos2008,Skorobogatiy2009}, the issue raised and the solution provided for electronics here are also highly relevant for seeking localized states of higher-order topological nature in phononic and photonic crystals \cite{Kim2020LSA,Wang2015PRL,Yang2015PRL,Liu2016NP,Slobozhanyuk2016NPh}, if following the same strategy.

\textit{System Model---.}
We employ the Slater-Koster multi-orbital tight-binding model to describe the monolayer graphene in a 16-dimensional Hilbert space: $ \mathcal{S} = \mathcal{S}_\text{sublatt.} \otimes \mathcal{S}_\text{orbit.} \otimes \mathcal{S}_\text{spin} $, where the sublattice, atomic orbital, and spin subspaces $ \mathcal{S}_\text{sublatt./orbit./spin} $ are spanned by the bases $ \mathcal{B}_\text{sublatt.} = \set{\ket{\text{A}}, \ket{\text{B}}} $, $ \mathcal{B}_\text{orbit.} = \set{\ket{s}, \ket{p_x}, \ket{p_y}, \ket{p_z}} $, and $ \mathcal{B}_\text{spin} = \set{\ket{\uparrow}, \ket{\downarrow}} $, respectively. By including the first-nearest-neighbor hoppings, the Hamiltonian reads
\begin{equation}
	H = H_\text{SK} + H_\text{SOC} + H_\text{M} + H_\text{AB}, \notag
\end{equation}
where $ H_\text{SK}, H_\text{SOC}, H_\text{M} $, and $ H_\text{AB} $ are terms resulting from wavefunction overlapping from Slater-Koster method \cite{Slater1954PR,Saito1998,Harrison1999,Zhong2017PRB}, atomic spin-orbit coupling (SOC), exchange field, and sublattice potential, respectively. These terms can be expressed in the second-quantization form
\begin{eqnarray}
	H_\text{SK} &=& \sum_{i\alpha} c_{i\alpha}^\dagger (\epsilon_{i\alpha} s_0) c_{i\alpha} + \sum_{\braket{ij}\alpha\beta} c_{i\alpha}^\dagger (t_{\alpha\beta} s_0)\ c_{j\beta} \label{eq:SK_hopping}, \\
	H_\text{SOC} &=& \xi_\text{SOC} \sum_{i,\alpha\beta} c_{i\alpha}^\dagger (\bm{s\cdot l})_{\alpha\beta}\ c_{i\beta} \label{eq:SOC}, \\
	H_\text{M} &=& M \sum_{i\alpha} c_{i\alpha}^\dagger (\bm{s\cdot\hat{n}}_M)\ c_{i\alpha} \label{eq:mag}, \\
	H_\text{AB} &=& U \sum_{i\in\text{A}, j\in\text{B}, \alpha} (c_{i\alpha}^\dagger c_{i\alpha} - c_{j\alpha}^\dagger c_{j\alpha}) \label{eq:sublat_pot},
\end{eqnarray}
where $ i $ and $ j $ label the atomic position in real space and $ \braket{\cdots} $ means to sum over nearest neighbors. $ \alpha $ and $ \beta $ take integer values from 0 to 3 and correspondingly stand for all the four outer-shell atomic orbitals in $ \mathcal{B}_\text{orbit.} $. The creation operator $ c_{i\alpha}^\dagger = (c_{i\alpha\uparrow}^\dagger, c_{i\alpha\downarrow}^\dagger) $ is understood with spin as its internal degree of freedom. The spin Pauli matrices are denoted as $ \bm{s} = (s_x, s_y, s_z) $. Hereinbelow, all energies are measured in eV and the length is in the unit of lattice constant $ a $, if not otherwise explicitly indicated.

\begin{figure*}[!htp]
  \centering
  \includegraphics[width=.95\textwidth]{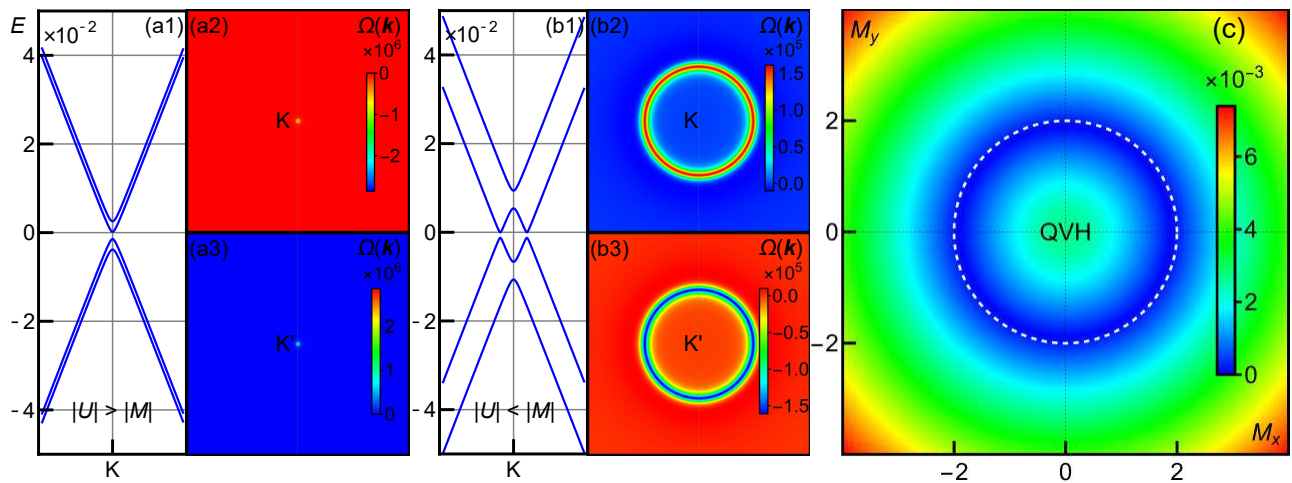}
  \caption{\label{fig1:bands_berrycurvatures_phases}
  (color online) (a1) Quantum valley Hall, $ (\mathcal{C}_\text{K}, \mathcal{C}_\text{K'}) = (-1, 1) $, with Berry curvature in (a2) and (a3). (b1) An insulating state, neither quantum anomalous Hall nor quantum valley Hall, with Berry curvature in (b2) and (b3). (c) Phase diagram with a circle phase border, with colors encoding the band gap. $ \xi_\text{SOC} = 0.100 $ ($ \Delta E_\text{bulk} = 0.00164 $) and energy in eV.}
\end{figure*}

Further detail of Eq.~(\ref{eq:SK_hopping}) can be found in Sec. SII in Supplementary Materials \cite{Zeng2021SM}. In Eq.~(\ref{eq:SOC}) the matrix element of the atomic spin-orbit coupling takes the form in the atomic-orbital subspace $ \mathcal{B}_\text{orbit.} $ as \cite{Liu2011PRB,Konschuh2010PRB}
\begin{equation}
	(\bm{s\cdot l}) = \text{i}
	\begin{pmatrix}
		0 & 0     & 0       & 0     \\
		0 & 0     & - g s_z & s_y   \\
		0 & g s_z & 0       & - s_x \\
		0 & - s_y & s_x     & 0
	\end{pmatrix},
\end{equation}
where $ g = 1 $ if not otherwise indicated. The parameter $ \xi_\text{SOC} $ is the atomic spin-orbit coupling strength and it is related to the ``intrinsic spin-orbit coupling" strength as $ t_2 = |\epsilon_s|\xi_\text{SOC}^2/(18 V_{sp\sigma}^2) $ \cite{Min2006PRB}. The latter considers a second-nearest-neighboring hopping among $ p_z $ orbitals and is responsible for a bulk gap $ \Delta E_\text{bulk} = 6\sqrt{3} t_2 $ \cite{Kane2005PRL}, which can be used as an estimation of the bulk gap in this work. The general form of magnetization term Eq.~(\ref{eq:mag}) has a unit vector $ \bm{\hat{n}}_M = (\sin\theta_M \cos\phi_M, \sin\theta_M \sin\phi_M, \cos\theta_M) $ specifying an arbitrary direction of the magnetization. In our consideration, we choose $ \theta_M = \pi/2 $ and $ \phi_M = 0 $, i.e., $ \bm{M} = M \hat{\textbf e}_x $, as it can be shown in the phase diagram Fig.~\ref{fig1:bands_berrycurvatures_phases}(c) that the in-plane direction of magnetization does not display its importance, concerning topological states.

\textit{Band Structure and Bulk Topology---.}	
Before we start to search for corner states, it is highly necessary to confirm the nontrivial topology of the bulk system. With an in-plane magnetization, the system can host two phases, as depicted in Fig.~\ref{fig1:bands_berrycurvatures_phases}. When the sublattice potential is larger ($ |U| > |M| $), a quantum valley Hall phase appears [Fig.~\ref{fig1:bands_berrycurvatures_phases}(a1)] with valley Chern numbers \cite{Berry1984PRSA,Qiao2010PRB,Jiang2012PRB} $ (\mathcal{C}_\text{K}, \mathcal{C}_\text{K'}) = (-1, 1) $ contributed by the Berry curvature in the proximity of the two valleys [Figs.~\ref{fig1:bands_berrycurvatures_phases}(a2) and (a3)]; whereas if $ |U| < |M| $, a ``VW"-shaped band structure taking forms around the global band gap and the Brillouin zone corners [Fig.~\ref{fig1:bands_berrycurvatures_phases}(b1)], however, the corresponding Chern number vanishes. In all the course, the in-plane direction of the magnetization is irrelevant, which can be drawn from Fig.~\ref{fig1:bands_berrycurvatures_phases}(c), where one can see an isotropic phase border, namely the white dashed circle.

\begin{figure}[htp!]
  \centering
  \includegraphics[width=.49\textwidth]{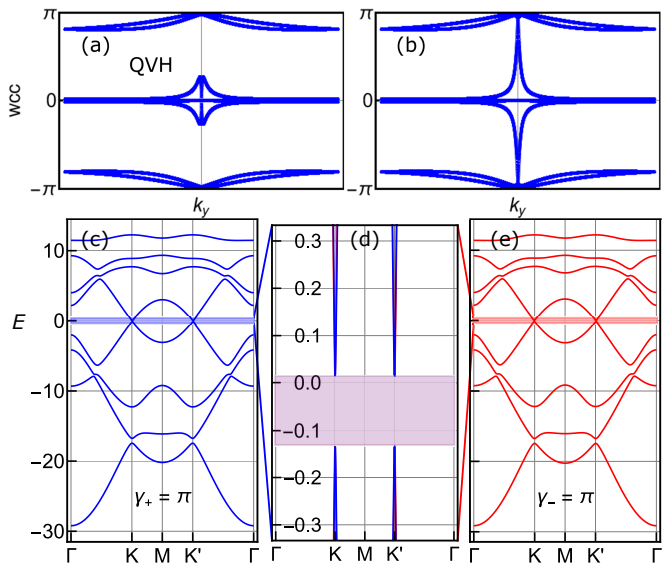}
  \caption{\label{fig2:bulk_topology} (color online) Characterization of the bulk topology. (a) and (b) Wannier charge center (wcc) evolution for the corresponding phases in Figs.~\ref{fig1:bands_berrycurvatures_phases}(a1) and (b1). (c) and (e) Band structures for $ H_\pm(k_y) $ in Eq.~(\ref{eq:H_+-}), with Zak phases $ \gamma_\pm = \pi $. (d) Zoom-in for the global gap.}
\end{figure}

However, the latter state does not have to be trivial just because of a vanishing Chern number. Without explicit indication, we set zero sublattice potential case hereafter. Then we can check the bulk topology with two methods: one is the bulk Wannier charge center (wcc) \cite{Yu2011PRB,Wang2019NJP} (see Sec. SIII in Supplementary Materials \cite{Zeng2021SM}) and the other is the mirror-graded Zak phase. The former generates the Wannier charge center evolution for the above quantum valley Hall effect and the other globally gapped state with a dominant in-plane magnetization in Figs.~\ref{fig2:bulk_topology}(a) and \ref{fig2:bulk_topology}(b), respectively. The major difference is that the latter is gapless but the former is not.

Furthermore, a mirror operator in the working representation $ (\mathcal{S}, \mathcal{B}) $ can be found to have this form \cite{Dresselhaus2008,Li2019}
\begin{equation}
  \mathcal{M}_x = \sigma_x \otimes \mathrm{diag}\{1, -1, 1, 1\} \otimes (\mathrm{i}s_x),
\end{equation}
whose determination is detailed in Sec. SIV of Supplementary Materials \cite{Zeng2021SM}, to commutate with the bulk Hamiltonian with a vanishing lattice momentum in the $ x $-direction: $ [\mathcal{M}_x, H(0, k_y)] = 0 $. The unitary operation $ \mathcal{U} $ that diagonalizes $ \mathcal{M}_x $ facilitates to direct-sum decompose $ H(0, k_y) $ as:
\begin{equation}\label{eq:H_+-}
  \mathcal{U} H(0, k_y) \mathcal{U}^\dagger = H_+(k_y) \oplus H_-(k_y).
\end{equation}
Both $ H_\pm(k_y) $'s spectra are globally gapped as shown in Figs.~\ref{fig2:bulk_topology}(c)-\ref{fig2:bulk_topology}(e), and they do not have much difference. Actually, every band is at least locally gapped from any one another. But neither of $ H_\pm(k_y) $ owns a chiral symmetry, hence the usual winding number evaluation technique \cite{Song2014PRB} is inapplicable here. Fortunately, we can find, by means of Wilson loop again, their corresponding Zak phases \cite{Sheng2019PRL,Zak1989PRL} and the result is $ \gamma_\pm = \pi $.

In summary, the gapless bulk Wannier charge center evolution and the nonvanishing mirror-graded Zak phase jointly confirm consistently that the gapped bulk phase in Fig.~\ref{fig1:bands_berrycurvatures_phases}(b1) is topologically nontrivial, but it is not a Chern insulator either. We then can reasonably expect it to be a second-order topological insulating phase. To find its higher-order embodiment, we next check corresponding systems with spatial dimension reduced.

\begin{figure}[htp!]
  \centering
  \includegraphics[width=.45\textwidth]{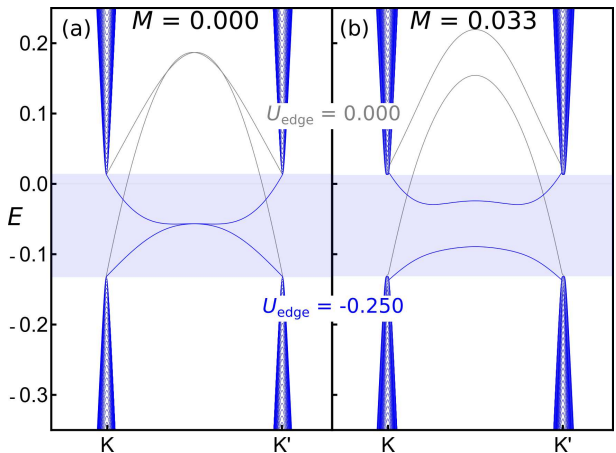}
  \caption{\label{fig3:zigzag_edge_modes} (color online) Band structures of zigzag nanoribbons with tunability of edge modes. (a) Quantum spin Hall effect. (b) Bulk insulating phase from in-plane magnetization. $ \xi_\text{SOC} = 1.000 $ ($ \Delta E_\text{bulk} = 0.164 $) and energy in eV.}
\end{figure}

\textit{In-gap Gapped Edge Modes with Tunability---.}
In the absence of magnetization, the system is in a $ \mathbb{Z}_2 $ quantum spin Hall effect, as the Kane-Mele model shows \cite{Kane2005PRL,Brey2020}. Indeed, our model gives a consistent result that the band structure of a zigzag nanoribbon is gapped out in bulk and in the meanwhile gapless helical edge modes link the valence and conduction bands [Fig.~\ref{fig3:zigzag_edge_modes}(a), gray curves]. However, a major difference emerges because of the absence of particle-hole symmetry that the intersection of the edge modes does not locate within the bulk gap (light-blue region), outside of which as well, consequently, is the edge gap opened up, when the in-plane magnetization takes effect [Fig.~\ref{fig3:zigzag_edge_modes}(b), gray curves]. This situation is detrimental to the quest of topological corner states locating within the edge gap, which is now able to rule out neither the bulk nor the edge states. By adiabatically introducing an onsite energy difference between the edge and bulk atoms \cite{Pantaleon2017PBCM,Pantaleon2018JPSJ}, the edge modes can be tuned to be inside the bulk gap, no matter the edge gap is absent or not [Figs.~\ref{fig3:zigzag_edge_modes}(a) and (b), blue curves]. The bulk band structures exhibit essentially no difference, before and after the introduction of such an onsite energy difference, as can be seen the blue and the gray bulk almost coincide with each other in Fig.~\ref{fig3:zigzag_edge_modes}. Because such a process is adiabatic, the same topological state maintains.

\begin{figure}[htp!]
  \centering
  \includegraphics[width=.49\textwidth]{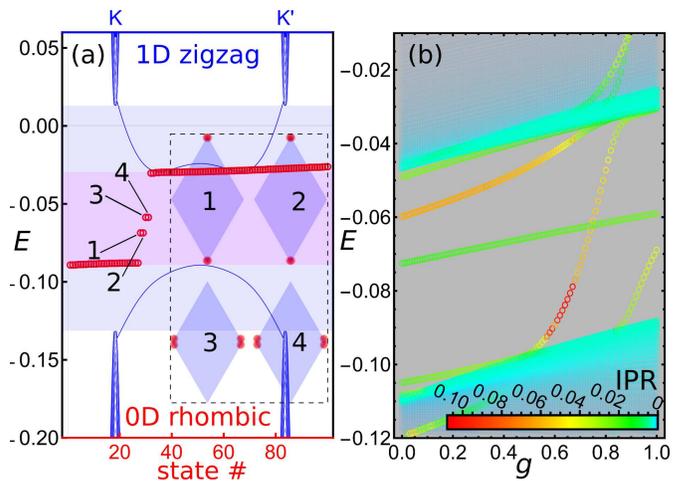}
  \caption{\label{fig4:0D_spectrum_and_corner_states}
  (color online) Topological corner states. (a) Blue band structure is the same as that in Fig.~\ref{fig3:zigzag_edge_modes}(b), red circles show the energy spectrum of a finite zigzag-edged rhombic sheet, the dash-framed inset shows the local density of states of corner states. (b) The energy spectrum of a same rhombic finite sheet with $ g \in [0, 1] $ in Eq.~(\ref{eq:SOC}). $ \xi_\text{SOC} = 1.000\ (\Delta E_\text{bulk} = 0.164), M = 0.033, U_\text{edge} = -0.250 $ and energy in eV.}
\end{figure}

\textit{Anomalous Topological Corner States---.}
Here we study corner states in a rhombic finite sheet edged with zigzag boundaries. In Fig.~\ref{fig4:0D_spectrum_and_corner_states}(a), the blue band structure exhibits the dispersion of the corresponding one-dimensional zigzag system, where a clear edge gap (light-red region) within the bulk gap (light-blue region) can be seen. Within this edge gap, four topological corner states appear in the spectrum of a finite rhombic sheet, as the red circles depict, where only a small portion of the states with low energies of interest is presented. Different from the situation reported previously, the four states come in pairs degenerated on non-zero energies. The lower-energy pair (labeled as 1 and 2) takes up the acute diagonal angles, and the higher-energy pair (3 and 4) the obtuse diagonal, as shown respectively in the dash-framed inset of Fig.~\ref{fig4:0D_spectrum_and_corner_states}(a). Every single topological corner state is a synthesis of both locality and nonlocality, and the four combined together leave no corners of the finite sheet unoccupied.

To reveal the newly emerged acute-angle occupying pair of corner states, we now vary the $ g $ factor in Eq.~(\ref{eq:SOC}) within the unit range $ [0, 1] $ to introduce the difference in SOC between the in-plane and the out-of-plane components. As shown in Fig.~\ref{fig4:0D_spectrum_and_corner_states}(b), the rightmost case with $ g = 1 $ is topological equality of that in Fig.~\ref{fig4:0D_spectrum_and_corner_states}(a), where the two pairs of corner states have relatively large inverse participation ratio ($ \text{IPR} = \sum_i |\braket{\psi|i}|^4 $ for a normalized state $ \ket{\psi} $ with $ \sum_i|\braket{\psi|i}|^2 = 1 $) \cite{Edwards1972JPCSSP,Murphy2011PRB,Calixto2015JSM,Misguich2016PRB,Tsukerman2017PRB} because of their comparatively high degree of localization and the acute-angle corner state is more localized than the obtuse-angle one. Furthermore, the spectrum evolves with a decreasing $ g $. The edge gap formed between the (green-blue) edge states moves downwards as a whole, in which course, the (green) obtuse-angle corner states remain in the middle of the gap, indicating the stability of the obtuse-angle corner state in two senses: (i) its energy position relative to the edge gap, and (ii) its degree of localization; both of which are effectively unaffected. On the other hand, the new acute-angle corner states do not enjoy those two kinds of robustness, because as $ g $ changes they change both its relative energy and inverse participation ratio. But in general the acute-angle corner state has a higher degree of localization and a better energy degeneracy. This shows the root of acute-angle corner state in the spin-orbit coupling between the in-plane $ p $-orbitals.

We further find that both types of corner states are robust against sheet shapes and orbital-dependent magnetization strengths, but the direction of magnetization can control the distribution of the acute-angle corner state. One can see that the two types of corner states are relatively stable because of their topological nature and meanwhile they differ from each other in their own right. More about the effects of sheet-shape and magnetization direction dependencies can be found in the last section of the Supplementary Materials \cite{Zeng2021SM}.

\textit{Summary and Discussion---.}
We propose the introduction of bulk-edge onsite energy contrast as a solution to the problem of identifying localized states with higher-order topology in two dimensions when the edge gap position acts as an obstacle. Through a combination of a series of evidence, we show that a second-order topological phase without particle-hole symmetry can be realized in monolayer graphene system with an in-plane magnetization, but its corner embodiment is not easily to find, hidden in the jungle of edge and bulk states. As of the nontrivial bulk topology, we show that the insulating state has a zero Chern number but the behavior of its Wannier charge center is gapless and both of the two mirror-graded subspaces carries a nonvanishing $ \pi $ Zak phase. When mentioning higher-order topology, the ``bulk-boundary correspondence" is an unavoidable topic, following which we check the zigzag-edged one- and zero-dimensional systems. In the former we find that, in the region of interest, a clear delineation cannot be achieved between the helical edge crossing and the bulk states. However, we also find that a bulk-edge onsite energy difference can amend it by tuning the edge crossing/gap into the bulk gap window. Then within that gap in gap, four midgap corner states with nonzero energy degeneracy in pairs can be found unambiguously. Furthermore, we examine the properties of the corner states and find that the obtuse-angle corner state is more stable against variation of coupling between $ p_x $ and $ p_y $ orbitals and the direction of in-plane magnetization. Bearing nontrivial topology in nature, both types are robust with respect to the shapes of finite sheets. The results are not limited to electronic system or two dimensions, and should also be of importance to higher-order topology in phononic and photonic crystals, due to the similar governing laws.

\textit{Acknowledgments---.} This work was finanically supported by the NNSFC (No. 11974327), Fundamental Research Funds for the Central Universities (WK3510000010, WK2030020032), Anhui Initiative in Quantum Information Technologies. We also thank the supercomputing service of AM-HPC and the Supercomputing Center of University of Science and Technology of China for providing the high performance computing resources.


\end{document}